\documentclass[a4paper,11pt]{article}

\usepackage{cite}
\usepackage{amsmath}
\usepackage{amscd}
\usepackage{amsfonts}
\usepackage{amssymb}

\begin{document}
\title{\bf{Gauge invariances of higher derivative Maxwell-Chern-Simons  field theory -- a new Hamiltonian approach}}
\author{
{\bf {\normalsize Pradip Mukherjee}}{\footnote{Also, Visiting Associate at S.~N.~Bose National Centre for Basic Sciences, JD Block, Sector III, Salt Lake, Kolkata-700098, India.}}\\
 {\normalsize \it Department of Physics, Barasat Government College,}\\
 {\normalsize \it Barasat,  West Bengal, India.}\\
  {\texttt{mukhpradip@gmail.com} \vspace{2ex}  }\\
 {\bf {\normalsize Biswajit Paul}}\\
 {\normalsize \it S.~N.~Bose National Centre for Basic Sciences,}\\
 {\normalsize \it Block-JD, Sector III, Salt Lake, Kolkata-700098, India.}\\
 {\texttt{bisu\_1729@bose.res.in}}
}
\vspace{-0.5truecm}

\date{}

\maketitle
\begin{abstract}
A new method of  abstracting the independent gauge invariances of higher derivative systems, recently introduced in \cite{bmp},  has been applied to higher derivative field theories. This has been discussed taking the extended Maxwell-Chern-Simons model as an example.  A new Hamiltonian analysis of the model is provided. This Hamiltonian analysis has been used to construct the independent gauge generator. An exact mapping between  the Hamiltonian gauge transformations and the U(1) symmetries  of the action has been established. 
\end{abstract}

\section {Introduction}

It is usual in field theories  to assume the Lagrangian  to be function of the fields and their first derivatives only. But there is no natural restriction which should confine us within this limitation. In fact, higher derivative theories were once thought to be attractive to get rid of infinities appearing in the scattering amplitudes \cite{podolsky1, podolsky2, podolsky3, mont, podolsky4}. However initial interest in such theories waned due to various difficulties in their formulation \cite{Pais} and also due to emergence of the powerful techniques of renormalisation. Notwithstanding this, research in the higher derivative theories still continued in a  steady, albeit slow, pace with discoveries of many interesting results. 
 Subsequently, new impetus to study higher derivative theories came from the attempt to quantize gravity \cite{smilga}. It is well known that the usual Einstein - Hilbert theory of gravity is not renormalizable because it contains dimensionfull coupling constant. By writing the gravity action in terms of the Weyl tensor we get a theory with dimensionless coupling constant which ensures renormalisability \cite{stelle,fradkin}. Such higher derivative theories  are now generically explored in terms of $f(R)$ gravity \cite{odnit1, soti, odnit2}. Higher derivative theories are again inevitable in the context of braneworld theory of quantum gravity. They have been obtained from string theory \cite{elie}, noncommutative theory \cite{clz}, and have been used in electrodynamics \cite{wf}, dark energy physics \cite{gib, caroll, woodard1}, inflation \cite{ani}, as ultra violet regulators\cite{slav, evens, bekeyv} and in other context\cite{lw1, lw2, cai, michel, polonyi}. Interesting connections of the higher derivative theories to non-commutative geometry and anyon physics are demonstrated \cite{pl5}.\\

 Like the usual theories the higher derivative theories may be endowed with gauge symmetry. In the canonical approach these are classified as singular theories. From the point of view of modern theoretical physics, gauge invariance is an essential component for physically interesting theories. Understanding the manifestations of the gauge symmetry in the canonical formalism has always been an issue of prime importance and has long been pursued in the literature\cite{D, HRT, rothe, sunder, heannux, gitman2}. There are several powerful techniques for abstracting the independent gauge transformations in the phase space and identifying them with the gauge invariances of the action\cite{cast, CGS, GP, BRR, BRR1, BRR2}. However all these works refer to usual first order theories. Though many investigations have been devoted to the Hamiltonian analysis of the higher derivative theories \cite{gitman2, ostro,  gitman1, pl1,pl2,pl3,pl4,N, B, morozov, dunin, AGMM, MARTINEZ}, certain important points remain unnoticed. One such issue is the extent of gauge degrees of freedom of a higher derivative theory.\\
 
  Indeed, the issue of gauge symmetry in higher derivative theory has its own peculiarities which demarcate  it from the usual theories.  For theories the Lagrangian of which contain the first derivative of the coordinates only, it has been proved quite generally that the number of independent gauge invariances of  a theory is equal to the number of independent primary first class constraint \cite{cast, BRR, BRR1}. This feature is, however not shared by the higher derivative theories. Consider for instance the example of a relativistic particle with rigidity \cite{P}\footnote{The model with the action (\ref{masparaction}) has been shown to be rich in physical content in a series of seminal papers by Plyuschay \cite{pl1, pl2}. The model has three different but related types of solutions, the massive, massless and tachyionic. On the other hand the $m=0$ analogue of (\ref{masparaction}) was proposed and analysed in \cite{pl3} where it was shown that the model is consistent classically only under the assumption that the velocity of the particle is greater than the velocity of light. This gauge dependent velocity arises due to a classical analogue of Zitterbewegung phenomenon for a massless spinning particle while the gauge invariant velocity is equal to the velocity of light. For $m \ne 0$ the parameter $\alpha$ multiplying the curvature term may take arbitrary values, whereas in the corresponding model with $m=0$ this parameter is quantized \cite{pl4}}. The action is given by:
 \begin{equation}
S=-m \int{\sqrt{-\dot{x}^{2}}}d\tau - \textbf{•}\alpha\int{\frac{\left( \left( \dot{x}\ddot{x}\right)^{2}-\dot{x}^{2}\ddot{x}^{2} \right)^\frac{1}{2}}{\dot{x}^{2}}}d\tau
\label{masparaction}
\end{equation}
A Hamiltonian analysis of the model \cite{N} exhibits that the Hamiltonian contains only a single arbitrary multiplier. Thus there is one independent gauge symmetry of the model. This is consistent with the fact that the action (\ref{masparaction}) has diffeomorphism invariance only. However, the number of independent primary first class constraints(PFC) of the theory is two. The number of independent PFCs  is thus more than the number of independent gauge degrees of freedom. The situation takes an interesting turn when the mass term is dropped from (\ref{masparaction}). Symmetries of the theory is now more general $W_{3}$- symmetry \cite{RR, RR1, RRO}. Here the number of independent gauge transformation is equal to the number of PFCs. These examples show that in the case of the higher derivative theories the number of independent gauge invariances sometimes matches with the number of PFCs and sometimes not. This fact was not much noticed and far less emphasized.\\

  Recently we have provided a general Hamiltonian method of abstracting the independent gauge transformations of the higher derivative theories \cite{bmp}. It is based on an equivalent first order formalism introduced earlier in the literature\cite{pl1, pl2,pl3}\footnote{for a related review on the subject see \cite{pl5}}. In this formalism the original higher derivative theory is converted to an equivalent first order theory by introducing new coordinates to account for the higher derivative terms. This leads to Lagrangian constraints which are imposed in the modified action by the Lagrange multiplier technique. These multipliers are then elevated to the level of independent fields. An unphysical sector is thus added in the phase space to proceed with the Hamiltonian analysis, followed by subsequent reduction to exhibit the physical sector. This equivalent first order formalism enables us to apply a structured algorithm  \cite{BRR, BRR1} for constructing the independent gauge generator of the first order theories, which has been applied to numerous models in the literature \cite{BMS1, BMS2, GHS, MS, samanta,BGMR,BGR}. The particular manner of extension of variables introduce novel connection in the phase space leading to new  restrictions on the gauge parameters. In case of the theory (\ref{masparaction}) we find that the new restrictions impose one more constraint on the gauge parameters leading to 2(nunber of independent PFC) - 1(number of new condition) = 1 independent gauge transformation. When the mass term is dropped the new restriction becomes trivial leading to two independent gauge transformation. Our method thus  clearly illustrate the inter relation of Hamiltonian gauge transformations with the PFCs for higher derivative theories, thereby explaining the apparent anomalies mentioned above. Also a general formulation for the construction of the Hamiltonian gauge generator containing the right number of independent gauge parameters is provided.\\
  
  The method advanced in \cite{bmp} offers a definite algorithm for abstracting independent gauge transformation for higher derivative theories in the canonical approach. In principle, it is applicable to both mechanical and field theoretic models. However so far this general method   has only been tested in the context of   particle models. A transition to field theories bring novel features even in the usual first derivative systems. It is thus natural to enquire how the method of \cite{bmp} works in the case of field theoretic models.   We would like to address this issue in the present paper.\\

 To illustrate the application of our method to field theories consider the action in $2+1$ dimension

\begin{equation}
S = \int d^{3}x  \left({ -\frac{1}{4}F_{\mu\nu}F^{\mu\nu} + \frac{g}{2} \epsilon^{ \alpha \beta \gamma }(\partial^{\rho}\partial_{\rho} A_{\alpha})(\partial_{\beta}A_{\gamma})}\right) 
\label{mcslag}
\end{equation}
where $F_{\mu \nu} =  \partial_{\mu}A_{\nu} - \partial_{\nu}A_{\mu}$. The second term in the action contains higher derivative terms and may be viewed as extension of the Chern-Simons piece. The theory (\ref{mcslag}) is thus called the extended Maxwell-Chern-Simons Model\cite{deser}. The choice of the model is dictated by the following:\\
\begin{enumerate}
\item{The model is a simple but an interesting field theoretic model \cite{deser}. It has been investigated several times in the recent past \cite{kumar, reyes}.}  
\item{Under the usual gauge transformation 
\begin{equation}
 A_{\mu} \rightarrow A_{\mu} + \partial_{\mu} \lambda 
 \label{gengaugeinv}
\end{equation}
the Lagrangian of (\ref{mcslag}) is invariant modulo total boundary terms. Thus the action (\ref{mcslag}) is invariant under (\ref{gengaugeinv}) if the function $ \lambda $ vanishes on the boundary. Thus the model offers a simple setting for comparing the Hamiltonian gauge symmetries with those of the action.}
\end{enumerate}

 Since our method is based on the equivalent first order approach, a detailed Hamiltonian analysis of the model (\ref{mcslag})  from the same point of view is required. Earlier Hamiltonian analysis of the model \cite{kumar, reyes} were based on Ostrogradski method \cite{ostro}. We need a constrained Hamiltonian analysis a la Dirac \cite{D} which we develop here. The constraint structure of the theory will be seen to have some nontrivial features which makes it interesting in it's own right. 

 Before concluding the introductory section let us elaborate the organisation of the paper. In Sec. 2 a review of the general method is discussed. Then in Sec 3  a detailed Hamiltonian analysis of the model (\ref{mcslag}) in the equivalent first order formalism is given. Note that this is a new calculation and distinct from that of \cite{kumar, reyes} which are based on the Ostrogradski method \cite{ostro} \footnote{In \cite{reyes} a gauge fixed version of (\ref{mcslag})  has been considered.}. In Sec. 4 the application of the method of \cite{bmp} to construct the Hamiltonian gauge generator is described. The gauge transformation generated by the gauge generator is compared with the transformation under (\ref{gengaugeinv}). Finally we conclude in Sec. 5. \\
\underline{} \section{General formalism -- a review}
   We begin with a general higher derivative theory given by the Lagrangian
\begin{equation}
L = L\left(x, \dot{x}, \ddot{x}, ..... , x^{\left(\nu\right)}\right)
\label{originallagrangean}
\end{equation}
where $x = x_n(n = 1,2,....,\nu)$ are the coordinates and $\dot{x}$ means derivative of $x$ with respect to time. $\nu$-th order derivative of time is denoted by $x^{\left(\nu\right)}$. 
The Hamiltonian formulation of the theory may be conveniently done by a variant of Ostrogradskii method. The crux of the method consists in embedding  the original higher derivative theory to an effective first order theory. We define the variables $q_{n,\alpha} \left(\alpha = 1, 2, ...., \nu - 1 \right)$ as
\begin{eqnarray}
q_{n,1}   &=& x_n\nonumber\\
q_{n,\alpha} &=& \dot{q}_{n,\alpha -1}, \left(\alpha > 1 \right)
\label{newvariables}
\end{eqnarray}
This leads to the following Lagrangian constraints
\begin{eqnarray}
q_{n,\alpha} - \dot{q}_{n,\alpha -1} = 0, \left(\alpha > 1 \right)
\label{lagrangeanconstraints}
\end{eqnarray}
which must be enforced by corresponding Lagrange multipliers .   
The auxiliary Lagrange function of this extended description
of the system is given by
\begin{eqnarray}
L^*(q_{n,\alpha},\dot{q}_{n,\alpha},\lambda_{n,\beta})
=L\left(q_{n,1},q_{n,2}\cdots,q_{n,\nu-1},
\dot{q}_{n,\nu-1}\right)+\sum_{\beta=2}^{\nu-1}\left(q_{n,\beta}-\dot{q}_{n,\beta-1}\right)\lambda_{n,\beta}\ ,
\label{extendedlagrangean}
\end{eqnarray}
where $\lambda_{n,\beta} (\beta = 2,\cdots , \nu - 1)$ are the Lagrange multipliers. If we consider these multipliers as independent fields then the Lagrangian $L^*$ becomes first order to which the well known methods of Hamiltonian analysis for  first order systems apply.
The momenta canonically
conjugate to the degrees of freedom $q_{n,\alpha}$,
$(\alpha=1,2,\cdots,\nu-1)$ and
$\lambda_{n,\beta}$ $(\beta = 2, \cdots,\nu-1)$
are defined, respectively, by,
\begin{equation}
p_{n,\alpha}=\frac{\partial L^{*}}{\partial \dot{q}_{n,\alpha}}\ ,\ \
\pi_{n,\beta}=\frac{\partial L^{*}}{\partial\dot{\lambda}_{n,\beta}}\ .
\end{equation}
These immediately lead at least to the following 
primary constraints,
\begin{equation}
\Phi_{n,\beta} \approx 0\ ,
\ \ \pi_{n,\beta} \approx 0\ ,\ \
\beta = 2,\cdots,\nu -1\ ,
\label{constraints}
\end{equation}
where
\begin{equation}
\Phi_{n,\beta}\equiv p_{n,\beta -1}+\lambda_{n,\beta}\ , \ \
\beta = 2,\cdots,\nu - 1\ .
\end{equation}
Note that depending on the situation whether the original Lagrangian $L$ is singular there may be more primary constraints. \\ 
   Assuming L to be singular the following possibilities may arise:
\begin{enumerate}

\item The original Lagrangian is singular but the additional constraints are all second class. Conserving the full set of primary constraints in time does not yield any secondary constraint. Rather,  all the multipliers in the total Hamiltonian will get fixed. The reduction of phase space may be done by implementing the second class constraints strongly provided we replace all the PBs by appropriate DBs.  

\item The original Lagrangian is singular and there are both primary second class and first class constraints among them. Conserving the primary constraints in time, secondary constraints will now be obtained. There may be both secondary second class and first class constraints. The second class constraints may be eliminated again by the DB technique. The first class constraints generate gauge transformations which are required to be further analysed. These constraints may yield further constraints and so on. The iterative process stops when no new constraints are generated. 
\end{enumerate}
From the point of view of gauge invariance the second case is important. Since the original Lagrangian system is replaced by the first order theory (\ref{extendedlagrangean}) the algorithm of \cite{BRR, BRR1} can be readily applied. All the first class constraints appear in the gauge generator G
\begin{equation}
G = \sum_a \epsilon^a \Phi_a
\label{217}
\end{equation}
where $\{\Phi_a\}$ is the whole set of (primary and secondary) first class constraints and $\epsilon^a$ are the gauge parameters. These parameters are however not independent. For a first order system the number of independent gauge parameters is equal to the number of independent PFCs. 
Following the algorithm of \cite{BRR, BRR1} we can express the dependent gauge parameters in terms of the independent set using the conditions
\begin{equation}
   \frac{d\epsilon^{a_2}}{dt}
 -\epsilon^{a}\left(V_{a}^{a_2}
+\lambda^{b_1}C_{b_1a}^{a_2}\right) = 0
\label{219}
\end{equation}
The indices $a_1, b_1 ...$ refer to the primary first class constraints while the indices $a_2, b_2 ...$ correspond to the secondary first class constraints.
The coefficients $V_{a}^{a_{1}}$ and $C_{b_1a}^{a_1}$ are the structure
functions of the involutive algebra, defined as\footnote{for theories with first class constraints only, {,} denotes Poisson bracket otherwise they refer to the appropriate Dirac bracket}
\begin{eqnarray}
\{H_{can},\Phi_{a}\} = V_{a}^b\Phi_{b}\nonumber\\
\{\Phi_{a},\Phi_{b}\} = C_{ab}^{c}\Phi_{c}
\label{2110}
\end{eqnarray}
and $\lambda^{a_1}$ are the Lagrange multipliers(associated with the primary first class constraints)  appearing in the expression of the total Hamiltonian.
Solving (\ref{219}) it is possible to choose $a_1$ independent
gauge parameters from the set $\epsilon^{a}$ and express $G$ of
(\ref{217}) entirely in terms of them. For the conventional first order theories this completes the picture. 
The situation for higher order theories is, however, different.
 This is because of the new constraints (\ref{lagrangeanconstraints}) appearing in the effective first order Lagrangian (\ref{extendedlagrangean}). Owing to these  we additionally require
\begin{eqnarray}
\delta q_{n,\alpha} - \frac{d}{dt}\delta{q}_{n,\alpha -1} = 0, \left(\alpha > 1 \right)
\label{varsgauge}
\end{eqnarray}
 These conditions may  reduce the number of independent gauge parameters further. Thus the number of independent gauge parameters is, in general, less than the number of primary first class constraints.
 \section{Hamiltonian analysis of the model in the equivalent first order formalism}
 In our approach the time derivative of the field $ A_{\mu} $ will be considered as additional fields. Thus it will be convenient to expand the Lagrangian of the model (\ref{mcslag})
in space and time parts. Using the mostly positive metric ($ \eta_{\mu\nu} $ = -, +, +) the Lagrangian is written as\\

 \begin{eqnarray}
 \nonumber
\mathcal{L} =  &=& \frac{1}{2} ( \dot{A}_{i}^{2} + (\partial_{i}A_{0})^{2} - (\partial_{i}A_{j})^{2} -2 \dot{A}_{i}\partial_{i}A_{0} + \partial_{i}A_{j}\partial_{j}A_{i} )+ \frac{g}{2} \epsilon_{ i j } ( -\ddot{A}_{0} +\nabla^{2} A_{0} )\partial_{i}A_{j} \\
&&- \frac{g}{2} \epsilon_{ i j } ( -\ddot{A}_{i} + \nabla^{2} A_{i} )\dot{A}_{j}+  \frac{g}{2} \epsilon_{ i j } ( -\ddot{A}_{i} + \nabla^{2} A_{i} )\partial_{j}A_{0} 
\label{mcslag1}
\end{eqnarray}
Here the fields are referred to their covariant components and dot represents derivative with respect to time. Note that the effect of the relativistic metric $\eta_{\mu\nu}$ has been taken care of explicitly in writing (\ref{mcslag1}). In the following subscripts from the middle of the Greek alphabet $ \mu $, $ \nu $ assume the values 0,1, 2 and those from the middle of the Latin alphabet $i$, $j$ take values 1 and 2. In any case, they just label the components and no further reference to the relativistic metric is implied.\\

 To analyse the model in the equivalent first order formalism we define the new coordinates
\begin{equation}
\xi_{1\mu} = A_{\mu}  \ \ \ {\rm{and}} \ \ \xi_{2\mu} = \dot{A}_{\mu}
\label{fielddef}
\end{equation}
This immediately imposes the constraint 
\begin{equation}
\xi_{2\mu} = \dot{\xi}_{1\mu} 
\label{lagconst}
\end{equation}
 The equivalent first order Lagrangian is obtained from (\ref{mcslag1}) using the definitions (\ref{fielddef}) as 

\begin{eqnarray}
\nonumber
\mathcal{L}^{\prime} &=& \frac{1}{2} ( \xi_{2i}\xi_{2i} + \partial_{i}\xi_{10}\partial_{i}\xi_{10} - \partial_{i}\xi_{1j}\partial_{i}\xi_{1j} -2 \xi_{2j}\partial_{j}\xi_{10} + \partial_{i}\xi_{1j}\partial_{j}\xi_{1i} )+ \frac{g}{2} \epsilon_{ i j } ( -\dot{\xi}_{20} +\nabla^{2} \xi_{10} )\partial_{i}\xi_{1j} \\
&&- \frac{g}{2} \epsilon_{ i j } ( -\dot{\xi}_{2i} + \nabla^{2} \xi_{1i} )\xi_{2j}+  \frac{g}{2} \epsilon_{ i j } ( -\dot{\xi}_{2i} + \nabla^{2} \xi_{1i} )\partial_{j}\xi_{10} + \xi_{0\mu} (\xi_{2\mu} - \dot{\xi}_{1\mu})
\label{auxilag}
\end{eqnarray}
Where the constraint (\ref{lagconst}) is enforced by the Lagrange multiplier $ \xi_{0\mu} $.
Henceforth $ \xi_{0\mu} $ will be considered as independent fields.\\

  To proceed with the canonical analysis we define the momenta $ \Pi_{0\mu} $,$ \Pi_{1\mu} $, $ \Pi_{2\mu} $ conjugate to the fields $ \xi_{0\mu} $, $ \xi_{1\mu} $, $ \xi_{2\mu} $ respectively in the usual way : 
\begin{eqnarray}
\Pi_{\alpha\mu} &=& \frac{\partial \mathcal{L}^{\prime}}{\partial \dot{\xi}_{\alpha\mu}} \ \ \ \ ; \ \ \ \ \ \ \ \ \ \ \  \alpha = 0, 1, 2
\end{eqnarray}  
    As a result the following primary constraints emerge. 
\begin{eqnarray}
\nonumber\\
\Phi_{0\mu} &=& \Pi_{0\mu}  \approx 0
\nonumber\\
\Phi_{1\mu} &=& \Pi_{1\mu} + \xi_{0\mu} \approx 0
\nonumber\\
\Phi_{20} &=& \Pi_{20} +  \frac{g}{2} \epsilon_{  i j }\partial_{i}\xi_{1j} \approx 0
\nonumber\\
\Phi_{2i} &=& \Pi_{2i} -  \frac{g}{2} \epsilon_{ i j }\xi_{2j} + \frac{g}{2} \epsilon_{ i j }\partial_{j}\xi_{10} \approx 0
\end{eqnarray}
The basic  Poisson brackets are 
\begin{eqnarray}
\nonumber\\
\lbrace{ \xi_{\alpha\mu}(\textbf{x}), \Pi_{\beta \nu}(\textbf{x}^{\prime}) }\rbrace &=& \delta_{\alpha\beta}\delta_{\mu\nu} \delta^{2}( \textbf{x} - \textbf{x}^{\prime} )
\label{basicpoisbarc}
\end{eqnarray}
where $ \alpha $, $ \beta $  = 0, 1, 2. This leads to the following  algebra of the primary constraint,
\begin{eqnarray}
\nonumber\\
\lbrace{ \Phi_{10}(\textbf{x}), \Phi_{2i}(\textbf{x}^{\prime}) }\rbrace &=&   -\frac{g}{2} \epsilon_{ij} \partial^{\prime }_{j}\delta^{2}( \textbf{x}-\textbf{x}^{\prime} )
\nonumber\\
\left\lbrace {\Phi_{1i}(\textbf{x}), \Phi_{20}(\textbf{x}^{\prime})}\right\rbrace &=& \frac{g}{2} \epsilon_{ij}\partial_{j}^{\prime} \delta^{2}(\textbf{x} - \textbf{x}^{\prime}) 
\nonumber\\
\lbrace{ \Phi_{2 i}(\textbf{x}), \Phi_{2j}(\textbf{x}^{\prime}) }\rbrace &=& - g \epsilon_{ i j}  \delta^{2}( \textbf{x}-\textbf{x}^{\prime} )
\nonumber\\
\lbrace{ \Phi_{0\mu}(\textbf{x}), \Phi_{1\nu}(\textbf{x}^{\prime}) }\rbrace &=& - \delta_{\mu\nu} \delta^{2}( \textbf{x}-\textbf{x}^{\prime} )
\label{constalgebra}
\end{eqnarray}
All other brackets between the constraints vanish. Apparently all the primary constraints have non trivial brackets among themselves. However, we can make the following linear combinations of the primary constraints 
\begin{eqnarray}
\nonumber
\Phi^{\prime}_{20} = \Phi_{20} + \frac{g}{2}\epsilon_{ij}\partial_{i}\Phi_{0j}  \approx 0
\nonumber\\
\Phi^{\prime}_{2i} = \Phi_{2i} + \frac{g}{2}\epsilon_{ij}\partial_{j}\Phi_{00}  \approx 0
\end{eqnarray}
Using the algebra of the primary constraints (\ref{constalgebra}) we find that the constraint algebra simplifies to   
\begin{eqnarray}
\nonumber\\
\lbrace{ \Phi_{0\mu}(\textbf{x}), \Phi_{1\nu}(\textbf{x}^{\prime}) }\rbrace &=& - \delta_{\mu\nu} \delta^{2}( \textbf{x}-\textbf{x}^{\prime} )
\nonumber\\
\lbrace{ \Phi_{2 i}^{\prime}(\textbf{x}), \Phi_{2j}^{\prime}(\textbf{x}^{\prime}) }\rbrace &=& - g \epsilon_{ i j}  \delta^{2}( \textbf{x}-\textbf{x}^{\prime} )
\end{eqnarray}
It will thus be convenient to replace  the original set of primary constraints  \{$ \Phi_{0\mu} $ , $ \Phi_{10} $, $ \Phi_{1i} $, $ \Phi_{20} $, $ \Phi_{2i} $\}  by \{$ \Phi_{0\mu} $, $ \Phi_{10} $, $ \Phi_{1i} $, $ \Phi_{20}^{\prime} $, $\Phi_{2i}^{\prime} $\}. 
Explicitly, the new set of primary constraints are 
\begin{eqnarray}
\nonumber\\
\Phi_{0\mu} &=& \Pi_{0\mu}  \approx 0
\nonumber\\
\Phi_{1\mu} &=& \Pi_{1\mu} + \xi_{0\mu} \approx 0
\nonumber\\
\Phi_{20}^{\prime} &=&  \Pi_{20}+ \frac{g}{2} \epsilon_{  i j }\partial_{i}\xi_{1j} + \frac{g}{2} \epsilon_{  i j }\partial_{i}\Pi_{0j} \approx 0
\nonumber\\
\Phi_{2i}^{\prime} &=& \Pi_{2i} -  \frac{g}{2} \epsilon_{ i j }\xi_{2j} + \frac{g}{2} \epsilon_{ i j }\partial_{j}\xi_{10} + \frac{g}{2} \epsilon_{  i j }\partial_{j}\Pi_{00} \approx 0
\label{primarycons}
\end{eqnarray}

 The canonical Hamiltonian is obtained by Legendre transformation as
\begin{eqnarray}
\nonumber\\
H_{can} &=& \int \mathcal{H}_{can} d^{2}\textbf{x}
\end{eqnarray}
Where $\mathcal{H}_{can}$ is the canonical Hamiltonian density, given by,
\begin{eqnarray}
\nonumber
\mathcal{H}_{can}&=& -\frac{1}{2} ( \xi_{2i}\xi_{2i} + \partial_{i}\xi_{10}\partial_{i}\xi_{10} - \partial_{i}\xi_{1j}\partial_{i}\xi_{1j} -2 \xi_{2j}\partial_{j}\xi_{10} +  \partial_{i}\xi_{1j}\partial_{j}\xi_{1i} )- \frac{g}{2} \epsilon_{ i j } \nabla^{2} \xi_{10} \partial_{i}\xi_{1j} \\
&&+ \frac{g}{2} \epsilon_{ i j } \nabla^{2} \xi_{1i} .\xi_{2j}- \frac{g}{2} \epsilon_{ i j } \nabla^{2} \xi_{1i} \partial_{j}\xi_{10} - \xi_{0\mu} \xi_{2\mu} 
\end{eqnarray}
The total Hamiltonian is
\begin{eqnarray}
H_{T} &=& \int d^{2}\textbf{x}( \mathcal{H}_{can} + \Lambda_{0\mu} \Phi_{0\mu} + \Lambda_{1\mu} \Phi_{1\mu} + \Lambda_{20} \Phi_{20}^{\prime} +\Lambda_{2i} \Phi_{2i}^{\prime}) 
\end{eqnarray}
The multipliers $ \Lambda_{0\mu} $, $ \Lambda_{1\mu} $, and $ \Lambda_{2\mu} $ are arbitrary at this stage.\\

The primary constraints (\ref{primarycons}) should be conserved in time i.e. there Poisson bracket with $ H_{T} $ should vanish. Conserving $ \Phi_{0\mu}$, $ \Phi_{1\mu}$, $ \Phi_{2i}^{\prime}$ in time the following multipliers are fixed, \\
\begin{eqnarray}
  \nonumber
  \Lambda_{00}  &=& \nabla^{2}\xi_{10} - \partial_{i}\xi_{2i} - g\epsilon_{ij}\nabla^{2}\partial_{i}\xi_{1j}
\nonumber \\
 \Lambda_{0i} &=& -\nabla^{2}\xi_{1i} + \partial_{i}\partial_{j}\xi_{1j} - g\epsilon_{ij}\nabla^{2}\partial_{j}\xi_{10} + \frac{g}{2}\epsilon_{ij}\xi_{2j}
 \nonumber\\
  \Lambda_{1\mu} &=& \xi_{2\mu}
 \nonumber\\
 \Lambda_{2i} &=& \frac{1}{2}( \nabla^{2}\xi_{1i} + \partial_{i}\xi_{20} ) + \frac{1}{g}\epsilon_{ij}( \partial_{j}\xi_{10} - \xi_{0j} - \xi_{2j} ) 
\label{multilpliers}
\end{eqnarray}
Only $ \Lambda_{20} $ remains arbitrary. Substituting these in the total Hamiltonian we find that it contains only one arbitrary multiplier $ \Lambda_{20} $. This shows that there is only one gauge degree of freedom, a result consistent with (\ref{gengaugeinv}).\\

 Conserving $ \Phi_{20}^{\prime} $ in time,  a secondary constraint emerges.
\begin{equation}
\Psi_{1} = \xi_{00} + \frac{g}{2}\epsilon_{ij}\partial_{i}\xi_{2j} \approx 0
\end{equation}
 From $ \dot{\Psi}_{1} = 0 $ we get 
 \begin{equation}
\left\lbrace{\Psi_{1}, H_{T}} \right\rbrace = 0
\end{equation} 
A straightforward calculation gives 
\begin{equation}
\Lambda_{00} - g \epsilon_{ij} \partial_{j} \Lambda_{2i} =0
\end{equation}
Using the values of $ \Lambda_{00} $ and $ \Lambda_{2i} $ from (\ref{multilpliers}) and simplifying we get
\begin{equation}
\partial_{i}\xi_{0i} - \frac{g}{2} \epsilon_{ij} \nabla^{2}\partial_{i}\xi_{1j} = 0
\end{equation}
which is a new secondary constraint
\begin{equation}
 \Psi_{2} = \partial_{i}\xi_{0i} - \frac{g}{2} \epsilon_{ij} \nabla^{2}\partial_{i}\xi_{1j} \approx 0
\end{equation}
The condition $ \left\lbrace {\Psi_{2}, H_{T}}\right\rbrace = 0 $ gives 
\begin{equation}
\partial_{i} \Lambda_{0i} - \frac{g}{2} \epsilon_{ij} \nabla^{2} \partial_{i} \Lambda_{1j} = 0
\end{equation}
Substituting the values of $ \Lambda_{0i}$ and $  \Lambda_{1j}$ the above equation reduces to the form 0 = 0. Hence the iterative process stops here giving no further constraints. The primary constraints of the theory are \{$ \Phi_{0\mu} $, $ \Phi_{1\mu} $, $ \Phi_{20}^{\prime} $, $ \Phi_{2i}^{\prime} $\} while the secondary constraints are $ \Psi_{1} $ and $ \Psi_{2} $.\\

 Using the Poisson brackets (\ref{basicpoisbarc}) the complete algebra of constraints can be worked out  as \\
\begin{eqnarray}
\nonumber\\
\left\lbrace{\Phi_{0\mu}(\textbf{x}), \Phi_{1\nu}(\textbf{x}^{\prime})} \right\rbrace &=& -\delta_{\mu \nu}\delta^{2}(\textbf{x}-\textbf{x}^{\prime})
\nonumber\\
\left\lbrace{\Phi_{2i}^{\prime}(\textbf{x}), \Phi_{2j}^{\prime}(\textbf{x}^{\prime})} \right\rbrace &=& -g \epsilon_{ij} \delta^{2}(\textbf{x}-\textbf{x}^{\prime})
\nonumber\\
\left\lbrace{\Psi_{1}(\textbf{x}), \Phi_{0\nu}(\textbf{x}^{\prime})} \right\rbrace &=& \delta_{0\nu }\delta^{2}(\textbf{x}-\textbf{x}^{\prime})
\nonumber\\
\left\lbrace{\Psi_{1}(\textbf{x}), \Phi_{2i}^{\prime}(\textbf{x}^{\prime})} \right\rbrace &=& -g \epsilon_{ij}\partial_{j}\delta^{2}(\textbf{x}-\textbf{x}^{\prime})
\nonumber\\
\left\lbrace{\Psi_{2}(\textbf{x}), \Phi_{0\mu}(\textbf{x}^{\prime})} \right\rbrace &=& \delta_{\mu i}\partial_{i}\delta^{2}(\textbf{x}-\textbf{x}^{\prime})
\nonumber\\
\left\lbrace{\Psi_{2}(\textbf{x}), \Phi_{1\mu}(\textbf{x}^{\prime})} \right\rbrace &=& -\frac{g}{2}\epsilon_{ij}    \delta_{j \mu} \nabla^{2}\partial_{i}\delta^{2}(\textbf{x}-\textbf{x}^{\prime})
\label{brackets3}
\end{eqnarray}
The constraint algebra appears to be complicated but new linear combinations will simplify the algebra. Before going into that discussion it is time to get rid of the unphysical variables $ \xi_{0\mu} $ and $ \Pi_{0\mu}  $.
\subsection{Calculation in reduced phase space}
The fields $ \xi_{0\mu} $ and  $ \Pi_{0\mu} $  can be eliminated by strongly imposing the constraints $\Phi_{0\mu}$ and $\Phi_{1\mu}$\footnote{Technically this should be done by replacing the  Poisson brackets by the corresponding Dirac brackets. However, the Dirac brackets here are trivial i.e. the Dirac brackets  between the remaining phase space variables are the same as the Poisson brackets .}. The remaining constraints of the theory can now be rewritten as 
\begin{eqnarray}
\nonumber\\
\Phi_{20} &=& \Pi_{20} +  \frac{g}{2} \epsilon_{  i j }\partial_{i}\xi_{1j}  \approx 0
\nonumber\\
\Phi_{2i} &=& \Pi_{2i} -  \frac{g}{2} \epsilon_{ i j }( \xi_{2j} - \partial_{j}\xi_{10} ) \approx 0
\nonumber\\
\Psi_{1} &=& -\Pi_{10} + \frac{g}{2}\epsilon_{ij}\partial_{i}\xi_{2j} \approx 0
\nonumber\\
\Psi_{2} &=& -\partial_{i}\Pi_{1i} - \frac{g}{2} \epsilon_{ij} \nabla^{2}\partial_{i}\xi_{1j} \approx 0
\end{eqnarray} 
The Poisson brackets between these constraints can be read from  (\ref{brackets3}). The nontrivial brackets are
\begin{eqnarray}
\nonumber\\
\left\lbrace{\Phi_{2i}(\textbf{x}), \Phi_{2j}(\textbf{x})} \right\rbrace &=& -g \epsilon_{ij}\delta^{2}(\textbf{x}-\textbf{x}^{\prime})
\nonumber\\
\left\lbrace{\Psi_{1}(\textbf{x}), \Phi_{2i}(\textbf{x}^{\prime})} \right\rbrace &=& -g \epsilon_{ij}\partial_{j}\delta^{2}(\textbf{x}-\textbf{x}^{\prime})
\end{eqnarray}
 We can form the linear combination  \\
\begin{equation}
\Psi_{1}^{\prime} = \Psi_{1} + \partial_{i}\Phi_{2i}
\end{equation}
It can be easily checked $\Psi_{1}^{\prime} $ has vanishing brackets with all other constraints. Replacing the set of constraints \{$ \Phi_{20} $, $ \Phi_{2i} $ ,$ \Psi_{1} $, and  $ \Psi_{2} $\} by the new set \{$ \Phi_{20} $, $ \Phi_{2i} $ , $ \Psi_{1}^{\prime} $, and $ \Psi_{2} $\} we find that there are three first class constraints $  \Phi_{20} $,  $ \Psi_{1}^{\prime} $, and $ \Psi_{2} $ and two second class constraints $ \Phi_{2i} $ . The classification of the constraints of the theory is tabulated in Table 1.\\
\begin{table}[h]
\label{table:constraints}
\caption{Classification of Constraints of the model (\ref{mcslag})}
\centering
\begin{tabular}{l  c  c}
\\[0.5ex]
\hline
\hline\\[-2ex]
& First class & Second class \\[0.5ex]
\hline\\[-2ex]
Primary &\ \ $\Phi_{20}$ &\ \ $\Phi_{2i}$\\[0.5ex]
\hline\\[-2ex]
Secondary &\ $\Psi_1^{\prime}$, $ \Psi_{2} $ &\ \ \\[0.5ex]
\hline
\hline
\end{tabular}
\end{table} 

Before proceeding further a degrees of freedom count will be instructive. The total number of phase space variables is 12. There are three first class constraints and two second class constraints. Hence the no of degrees of freedom is
\begin{equation}
\nonumber
12- (2 \times 3 + 2) = 4 
\end{equation}
We find that the number of degrees of freedom is doubled compared with the Maxwell theory, which is expected due to the higher derivative nature \cite{ostro}. 
\subsection{Reduction of second class constraints $ \Phi_{2i} $}
After the elimination of the unphysical sector ( $ \xi_{0\mu} $, $ \Pi_{0\mu} $ ), the total Hamiltonian becomes \\
\begin{equation}
H_{T}(\textbf{x}) = \int d^{2}\textbf{x} ( \mathcal{H}_{can}(\textbf{x}) + \Lambda_{20}(\textbf{x}) \Phi_{20}(\textbf{x}) + \Lambda_{2i}(\textbf{x}) \Phi_{2i}(\textbf{x}))
\end{equation}
Where $ \mathcal{H}_{can} $ is the canonical Hamiltonian density given by
\begin{eqnarray}
\nonumber
\mathcal{H}_{can} &=& -\frac{1}{2} ( \xi_{2i}\xi_{2i} + \partial_{i}\xi_{10}\partial_{i}\xi_{10} - \partial_{i}\xi_{1j}\partial_{i}\xi_{1j} -2 \xi_{2j}\partial_{j}\xi_{10} +  \partial_{i}\xi_{1j}\partial_{j}\xi_{1i} )- \frac{g}{2} \epsilon_{ i j } \nabla^{2} \xi_{10} \partial_{i}\xi_{1j} \\
 && + \frac{g}{2} \epsilon_{ i j } \nabla^{2} \xi_{1i} ( \xi_{2j} - \partial_{j}\xi_{10}) + \Pi_{1\mu} \xi_{2\mu} 
\end{eqnarray}
 And
 \begin{equation}
 \nonumber
 \Lambda_{2i} = \frac{1}{2}( \nabla^{2}\xi_{1i} + \partial_{i}\xi_{20} ) + \frac{1}{g}\epsilon_{ij}( \partial_{j}\xi_{10}+ \Pi_{1j}  - \xi_{2j} )  
  \end{equation}
$ \Lambda_{20} $ is arbitrary. It signifies that there is one continuous gauge degree of freedom. \\

	In the next section we will explicitly construct the gauge generator using the method given in \cite{bmp}. Since the method is directly applicable to theories with first class constraint only, we have to eliminate the second class constraints of our theory. Following Dirac's method of constraint Hamiltonian analysis we can strongly put the second class constraints to be zero if the Poisson brackets are replaced by the corresponding Dirac brackets.\\
	
	  The Dirac bracket between two phase space variables A and B is defined by
  \begin{equation}
\left[ {A(\textbf{x}), B(\textbf{x}^{\prime})}\right]   = \left\lbrace {A(\textbf{x}), B(\textbf{x}^{\prime})}\right\rbrace - \int \left\lbrace {A(\textbf{x}), \Phi_{2i}(\textbf{y})} \right\rbrace \Delta_{ij}^{-1}(\textbf{y},\textbf{z})\left\lbrace {\Phi_{2j}(\textbf{z}), B(\textbf{x}^{\prime}) } \right\rbrace  d^{2}\textbf{y} d^{2}\textbf{z}  
  \end{equation}
Where   $\Delta_{ij}^{-1}(\textbf{x} , \textbf{x}^{\prime})$ is the inverse of the matrix 
\begin{equation}
\Delta_{ij}(\textbf{x} , \textbf{x}^{\prime}) = \lbrace{ \Phi_{2 i}(\textbf{x}), \Phi_{2j}(\textbf{x}^{\prime}) }\rbrace 
\end{equation} 
The nontrivial Dirac brackets between the phase space variables are calculated as
\begin{eqnarray}
\nonumber
\left[ {\xi_{1\mu}(\textbf{x}), \Pi_{1\nu}(\textbf{x}^{\prime})}\right] &=& \delta_{\mu\nu} \delta^{2}(\textbf{x}-\textbf{x}^{\prime})
\nonumber\\
\left[ {\xi_{2i}(\textbf{x}), \xi_{2j}(\textbf{x}^{\prime})}\right] &=& \frac{1}{g} \epsilon_{ij} \delta^{2}(\textbf{x}-\textbf{x}^{\prime})
\nonumber\\
\left[ {\xi_{2i}(\textbf{x}), \Pi_{2j}(\textbf{x}^{\prime})}\right] &=& \frac{1}{2} \delta_{ij} \delta^{2}(\textbf{x}-\textbf{x}^{\prime})
\nonumber\\
\left[ {\xi_{2i}(\textbf{x}), \Pi_{10}(\textbf{x}^{\prime})}\right] &=& -\frac{1}{2} \partial^{\prime}_{i} \delta^{2}(\textbf{x}-\textbf{x}^{\prime})
\nonumber\\
\left[ {\Pi_{2i}(\textbf{x}), \Pi_{10}(\textbf{x}^{\prime})}\right] &=& \frac{g}{4} \epsilon_{ij} \partial^{\prime}_{j} \delta^{2}(\textbf{x}-\textbf{x}^{\prime})
\nonumber\\
\left[ {\Pi_{2i}(\textbf{x}), \Pi_{2j}(\textbf{x}^{\prime})}\right] &=& \frac{g}{4} \epsilon_{ij} \delta^{2}(\textbf{x}-\textbf{x}^{\prime})
\label{basicdirac}
\end{eqnarray}
All other Dirac brackets are the same as the corresponding Poisson brackets.
\section{Construction of the Gauge generator}
As has been mentioned earlier we will follow the method of \cite{bmp} to construct the gauge generator containing the exact number of independent gauge parameters. The essence of the method has been reviewed in Sec. 2. Accordingly, we rename the constraints as $ \Omega_{1}=\Phi_{20} $, $ \Omega_{2}=\Psi_{1}^{\prime} $ and $ \Omega_{3}=\Psi_{2} $. The gauge generator is \\
\begin{equation}
G = \int\epsilon_{a} \Omega_{a} d^{2}\textbf{x}
\end{equation}
which is a field theoretic extension of (\ref{217}).  These structure functions are now defined by  
\begin{eqnarray}
\nonumber
\left[{H_{can}, \Omega_{a}(\textbf{x})} \right] &=& \int d^{2}\textbf{y}  V_{ab}(\textbf{y}, \textbf{x}) \Omega_{b}(\textbf{y})
\nonumber\\
\left[{\Omega_{a}(\textbf{x}), \Omega_{b}(\textbf{y})} \right] &=& \int d^{2}\textbf{z}  C_{abc}(\textbf{z}, \textbf{x}, \textbf{y}) \Omega_{c}(\textbf{z})
\label{fieldstruc} 
\end{eqnarray}
 and the master equation (\ref{219}) takes the form 
\begin{eqnarray}
0 = \frac{d\epsilon_{a_{1}}(\textbf{x})}{dt} - \int d^{2}\textbf{y} \epsilon_{b}(\textbf{y}) V_{ba_{1}}(\textbf{x}, \textbf{y}) - \int d^{2}\textbf{y} d^{2}\textbf{z} \epsilon_{b}(\textbf{y}) \Lambda_{c_{1}}(\textbf{z}) C_{c_{1}ba_{1}}(\textbf{z}, \textbf{y}, \textbf{x}) 
\label{mastereqn3}
\end{eqnarray}
Note that Dirac brackets appear on the left hand sides of equations (\ref{fieldstruc}). This is because there were second class constraints in our theory which have been eliminated by the Dirac bracket formalism.\\

 Using the defining relations (\ref{fieldstruc}) and the Dirac brackets (\ref{basicdirac}) we find that the only nonvanishing $V_{ab}$ are given by 
\begin{eqnarray}
\nonumber
V_{12}(\textbf{x}, \textbf{y}) &=& - \delta^{2}(\textbf{x}-\textbf{y})
\nonumber\\
V_{23}({\textbf{x}, \textbf{y}}) &=& - \delta^{2}(\textbf{x}-\textbf{y})
\end{eqnarray}
Similarly from the algebra of the constraints  we find all $C_{abc}=0$. Substituting these values in the equation (\ref{mastereqn3}) we get the following conditions on the gauge parameters $\epsilon_{a}$
\begin{eqnarray}
\nonumber
\dot{\epsilon}_{2} + \epsilon_{1} &=& 0
\nonumber\\
\dot{\epsilon_{3}} + \epsilon_{2} &=& 0
\end{eqnarray}
Solving these we find 
\begin{eqnarray}
\nonumber
  \epsilon_{1} &=& \ddot{\epsilon}_{3}
\nonumber\\
  \epsilon_{2} &=& -\dot{\epsilon_{3}}
\end{eqnarray}
Hence the desired gauge generator assumes the form
\begin{equation}
G = \int d^{2}{x} (\ddot{\epsilon}_{3} \Omega_{1} - \dot{\epsilon}_{3} \Omega_{2} +\epsilon_{3}\Omega_{3})
\label{generator3}
\end{equation}
It is immediately observed that G contains one arbitrary gauge parameter namely $ \epsilon_{3} $.\\

 We still have the additional restrictions (\ref{varsgauge}). In our case this leads to the condition 
 \begin{equation}
\delta\xi_{2\mu} = \frac{d}{dt} \delta \xi_{1\mu}
\label{addrestriction2}
\end{equation} 
where $  \delta \xi_{1\mu} $, $  \delta \xi_{2\mu} $ are the gauge variations of $ \xi_{1\mu} $ and $ \xi_{2\mu} $ respectively. Using the generator G (\ref{generator3}) we get
\begin{equation}
\delta \xi_{2\mu} = \left\lbrace {\xi_{2\mu}, G}\right\rbrace   =  \partial_{\mu} \dot{\epsilon}_{3}
\end{equation}
Similarly 
\begin{equation}
\delta \xi_{1\mu} = \left\lbrace {\xi_{1\mu}, G}\right\rbrace   =  \partial_{\mu} \epsilon_{3}
\label{gaugevarxi1}
\end{equation}
Clearly the additional restriction (\ref{addrestriction2}) is identically satisfied. Thus no more restriction is imposed on the gauge parameters. \\

 Finally we look at the comparison of the transformations generated by the Hamiltonian gauge generator with Lagrangian gauge symmetry  (\ref{gengaugeinv}). Since $ \xi_{1\mu} = A_{\mu} $ we have
 \begin{equation}
\delta A_{\mu} = \partial_{\mu} \epsilon_{3}
\end{equation}
from (\ref{gaugevarxi1}). This is the same transformation as (\ref{gengaugeinv}) if we put $ \epsilon_{3} $ = $ \lambda $.
\section{Conclusion}
 Higher derivative  systems were once invoked in field theory to account for the ultraviolet divergences
   \cite{ podolsky1, podolsky2, podolsky3, mont, podolsky4}. Later the initiative was stalled partly because of various difficulties in their formulation \cite{Pais} and also due to emergence of the powerful techniques of renormalisation. In recent times interest in the higher derivative field  theories has been rejuvenated due to there relevance in quantum gravity \cite{stelle, fradkin, odnit1, soti, odnit2}. In this context understanding the gauge invariances of these theories from the canonical approach becomes an urgent problem. Though there are a number of Hamiltonian analysis of higher derivative theories available in the literature \cite{ostro,  gitman1, pl1,  N, B, gitman2,  morozov, dunin, AGMM, MARTINEZ}, certain important issues have been overlooked. One such issue is the abstraction of the independent gauge degrees of freedom. Indeed some confusion regarding this is evident. In the usual first order theories we can prove in general that the number of independent parameters in the Hamiltonian gauge generator is equal to the number of independent primary first class constraints(PFCs) of the theory \cite{cast, BRR, BRR1}. This connection seems to be violated in the case of the higher derivative theories \cite{bmp}. Thus in the Hamiltonian analysis of the relativistic particle model with curvature \cite{N} one observes two independent primary first class constraints though the number of gauge degrees of freedom is only one. The problem of  gauge invariances in higher derivative theories contains peculiar surprises. If the action of the relativistic particle is given by the curvature term only\footnote{this model was introduced and its physical content clarified in \cite{pl3, pl4}} the gauge transformations are found to satisfy the $ W_{3} $ algebra \cite{RR, RR1, RRO}. The independent gauge degrees of freedom is two which is equal to the number of independent PFCs. Thus there seems to be no regular connection between the number of independent gauge transformations with the number of independent PFCs for the higher derivative systems. 
 A general approach of constructing the Hamiltonian gauge generator of higher derivative systems have very recently been proposed which clearly explains this apparent anomaly \cite{bmp}. It also provides a general method of constructing the gauge generator containing the right number of independent gauge parameters. The method is sufficiently general so as to be applicable to both mechanical and field theoretic model. However so far the method is applied to particle models only. In this paper we have for the first time applied the formalism developed in  \cite{bmp} to field theories taking the extended Maxwell-Chern-Simons(M-C-S) model as example.\\
 
  The extended M-C-S model is a simple but  interesting example of higher derivative field theory and has been investigated many times in the recent past \cite{kumar, reyes}. The Lagrangian gauge symmetry of the model is the obvious U(1) gauge symmetry. The model thus provides a benchmark for the comparison of the Hamiltonian and Lagrangian gauge symmetries.  Since the method of \cite{bmp} is based on an equivalent first order formalism we have given a detailed Hamiltonian analysis of the model from that approach. Note that this is  a new calculation different from the earlier Hamiltonian analysis of the model \cite{kumar, reyes}.  This Hamiltonian analysis was then used to construct the independent gauge generator. Correspondence of the transformation generated by this  has been established with the gauge symmetries of the action and an exact mapping was demonstrated between the Lagrangian and Hamiltonian gauge parameters.\\
  
  Though illustrated by a simple example, our analysis given in this paper provides a facility to analyze the independent gauge invariances of  more intricate higher derivative models. From the connections of higher derivative theories with such modern contexts of anyon physics and non-commutative geometry \cite{pl5} and the relevance of higher derivative theories in the modern theories of gravity \cite{soti} this facility will indeed be welcome.

 \section*{Acknowledgement}
P.M. and B.P. would like to thank R. Banerjee for useful discussion. P. M. acknowledges the facilities provided by IUCAA and SNBNCBS where parts of the work were done. B.P. thanks  CSIR for financial support.


\begin{thebibliography}{999}
\bibitem{bmp} R. ~Banerjee, P. ~Mukherjee, B. ~Paul, JHEP 1108:085,2011. arXiv : 1012.2969 .
\bibitem{podolsky1} B.~Podolsky, Phys.\ Rev.\  {\bf 62}, 68 (1942).
\bibitem{podolsky2} B.~Podolsky and C.~Kikuchi,
Phys.\ Rev.\  {\bf 65}, 228 (1944).
\bibitem{podolsky3} B.~Podolsky and C.~Kikuchi, Phys.\ Rev.\
{\bf 67}, 184 (1945).
\bibitem{mont} D.~J.~Montgomery, Phys.\ Rev.\  {\bf 69}, 117 (1946)
\bibitem{podolsky4}B.~Podolsky and P.~Schwed, Rev.\ Mod.\ Phys.\  {\bf 20}, 40 (1948).
\bibitem{Pais} A.~Pais and G.~E.~Uhlenbeck, Phys.\ Rev.\  {\bf 79}, 145 (1950).



 

 
\bibitem{smilga}A. ~V. ~Smilga, Nuclear Physics B \textbf{706 }(2005) 598614.
\bibitem{stelle} K.~S. ~Stelle,  Phys. Rev.  {\bfseries D16} (1977), 953.
\bibitem{fradkin} E.~S. ~Fradkin, A.A. Tseytlin, Nucl. Phys. {\bfseries B201} (1982), 469.
\bibitem{odnit1} S. Nojiri, S.D. Odintsov, Int.J.Geom.Meth.Mod.Phys.\textbf{4}(2007), 115; arXiv:hep-th/0601213v5.
\bibitem{soti} T.P. Sotiriou,Rev. Mod. Phys. 82, 451-497 (2010); arXiv:0805.1726.
\bibitem{odnit2} S. ~Nojiri, S. ~D. ~Odintsov, 	Phys.Rept.\textbf{505}(2011),59.
\bibitem{elie} D. A. Eliezer and R. P. Woodard, Nucl.\ Phys.\ B \textbf{325}, 389 (1989).
\bibitem{clz} C.~S.~Chu, J.~Lukierski and W.~J.~Zakrzewski, Nucl.\ Phys.\  B {\bf 632}, 219 (2002).
  \bibitem{wf} J.~A.~Wheeler and R.~P.~Feynman, Rev.\ Mod.\ Phys.\  {\bf 21}, 425 (1949).
  \bibitem{gib} G.~W.~Gibbons; arXiv:hep-th/0302199.
\bibitem{caroll} S.~M.~Carroll, M.~Hoffman and M.~Trodden, Phys.\ Rev.\  D {\bf 68}, 023509 (2003).
\bibitem{woodard1}R.~P.~Woodard, Lect.\ Notes Phys.\  {\bf 720}, 403 (2007).
\bibitem{ani} A.~Anisimov, E.~Babichev and A.~Vikman,  JCAP {\bf 0506}, 006 (2005).
\bibitem{slav} A.~A.~Slavnov, Teor.\ Mat.\ Fiz.\  {\bf 13}, 174 (1972).
\bibitem{evens} D.~Evens,J.~W.~Moffat, G.~Kleppe and R.~P.~Woodard, Phys.\ Rev.\  D {\bf 43}, 499 (1991).
\bibitem{bekeyv} T.~D.~Bakeyev and A.~A.~Slavnov, Mod.\ Phys.\ Lett.\  A {\bf 11}, 1539 (1996).
\bibitem{lw1}T. D. Lee and G. C. Wick,  Nucl. Phys.B 9, 209 (1969).
\bibitem{lw2}T. D. Lee and G. C. Wick,  Phys. Rev. D 2, 033 (1970).
\bibitem{michel} Y. ~Michel, B. ~Pioline JHEP \textbf{2007}, 103 (2007).
\bibitem{cai}Y. ~Cai and D. ~A. ~Easson, Journal of Cosmology and Astroparticle Physics 2010, 002 (2010)
\bibitem{polonyi} J. ~Polonyi and A. ~Siwek, Phys. Rev. D \textbf{84}, 085014 (2011).
\bibitem{pl5}M.~S. ~Plyuschay, J. Theor. Phys. 3N10:17-31, 2006, [arXiv: math-ph/0604022]. 
\bibitem{D} P.A.M. Dirac, Can. J. Math. {\bf 2} (1950) 129;{\it Lectures on Quantum Mechanics}, Yeshiva University, 1964.
\bibitem{HRT} A.~Hanson, T.~Regge, C.~Tietelboim, ``\textit {Constrained Hamiltonian System}'', (Accademia Nazionale Dei Lincei, Roma, 1976).
\bibitem{rothe}  H. J. Rothe, K. D. Rothe, ``\textit{Classical And Quantum Dynamics of Constrained Hamiltonian Systems}'' .World Scientific  Lecture Notes in Physics - Vol. 81.
\bibitem {sunder} K.~Sundermeyer, ``\textit {Lecture Notes in Physics 169, Constrained Dynamics}'', (Springer-Verlag, 1982).
 \bibitem{heannux} M. ~Henneaux, C. ~Teitelboim, ``\textit{ Quantization of Gauge Systems}'', Princeton University Press.
 \bibitem{gitman2}  D. M.~Gitman and I. V.~Tyutin, `` \textit{Quantization of Fields with Constraints}'',
{\it  Springer -- Verlag, Berlin, Heidelberg (1990) 291 p}.
\bibitem{cast} L.~Castellani, Ann.~Phys.~{\bf{143}}, (1982) 357.
\bibitem{CGS} M.E.V. Costa, H.O. Girotti and T.J.M. Simoes, Phys. Rev. {\bf D32} (1985) 405.
 
 
\bibitem{GP} X. Gracia and J. M. Pons, J. Phys. A:Math. Gen. {\bf{28}} (1995), 7181.

\bibitem{BRR} R.~Banerjee, H.~J.~Rothe and K.~D.~Rothe, Phys.~Lett.~{\bf{B 463}} (1999) 248; arXiv : hep-th/9906072.
\bibitem{BRR1}R.~Banerjee, H.~J.~Rothe and K.~D.~Rothe, Phys.~Lett.~{\bf{B 479}} (2000) 429 arXiv : hep-th/9907217.
\bibitem{BRR2} R. ~Banerjee, H.~J. ~Rothe, K.~D. ~Rothe,  J.Phys.\textbf{A33}(2000)2059;  arXiv : hep-th/9909039. 
\bibitem{ostro} M. Ostrogradsky, \textit{Mem. Ac. St. Petersbourg} {\bf V14} (1850) 385.
\bibitem{gitman1} D. M.~Gitman, S. L.~Lyakovich and I. V.~Tyutin, Sov. Phys. Journ., {\bf{26}}, 730 (1983).
 \bibitem{pl1} M.~S. ~Plyuschay,  Mod. Phys. Lett. {\bfseries A3} (1988), 1299. 
\bibitem{pl2}M.~S. ~Plyuschay, Int. J. Mod. Phys. A4: 3851-3865 (1989).
\bibitem{pl3}M.~S. ~Plyuschay, Mod. Phys. Lett. A4: 837-847, (1989).
\bibitem{pl4}M.~S. ~Plyuschay, Phys. Lett. B 243; 383-388, 1990.

  \bibitem{N} V.V.Nesterenko,J. Phys. {\bf{A22}} (1989) 1673.
  \bibitem{B} I. L. Buchbinder, S. L. Lyakhovich, V. A. Krykhtin, Class. Quant. Grav. \textbf{10}(1993)2083.
  
 \bibitem {morozov} A. Morozov,Theoretical and Mathematical Physics, \textbf{157(2)}(2008): 1542,  arXiv:0712.0946.
 \bibitem{dunin} P. Dunin -- Barkowski and A. Sleptsov,Theoretical and Mathematical Physics, \textbf{158(1)}(2009): 61; arXiv : 0801.4293. 
\bibitem{AGMM} K. Andrzejewski, J. Gonera, P. Machalski, P. Maslanka,Physical Review \textbf{D 82}(2010) , 045008; arXiv : 1005.3941.
\bibitem{MARTINEZ} Pedro D. Prieto-Martinez, Narciso Roman-Roy, J.Phys.A4 4:\textbf{385203},(2011). arXiv 1106.3261 .
 

 \bibitem{P} R. D. Pisarski, Phys. Rev. D {\bf{34}} (1986), 670.

\bibitem{RR} E. Ramos and J. Roca, Nucl. Phys. {\bf{B436}} (1995), 529.
\bibitem{RR1} Nucl.Phys. {\bf{B452}} (1995), 705.
  \bibitem{RRO} E. Ramos and J. Roca, Phys.Lett. B\textbf{366} (1996) 113; arXiv:hep-th/9506088.
  
   



\bibitem{BMS1}R.~Banerjee, P.~Mukherjee, A.~Saha, Phys.~Rev.~{\bf{D 70}} (2004) 026006, arXiv : hep-th/0403065.
\bibitem{BMS2}R.~Banerjee, P.~Mukherjee, A.~Saha, Phys.~Rev.~{\bf{D 72}} (2005) 066015; arXiv : hep-th/0501030 .
\bibitem{GHS}S.~Gangopadhyay, A.~Ghosh~Hazra, A.~Saha, Phys.~Rev.~{\bf {D 74}} (2006) 125023,arXiv : hep-th/0701012.
\bibitem{MS} P. Mukherjee and A. Saha, Int. J. Mod. Phys. A24, 4305 (2009).
\bibitem{samanta} S. ~Samanta, Int.J.Theor.Phys.\textbf{48}(2009)1436;  arXiv: 0708.3300.
\bibitem{BGMR} R. Banerjee, S. Gangopadhyay, P. Mukherjee, D. Roy, JHEP 1002:075, (2010);  arXiv  : 0912.1472.
\bibitem{BGR} R. ~Banerjee, S. ~Gangopadhyay, D. ~Roy; arXiv : 1108.4591v3.
\bibitem{deser} S. Deser, R. Jackiw, Physics Letters B \textbf{451} (1999)73.
\bibitem{kumar} S. ~Kumar,  Int. J. Mod. Phys. A \textbf{18}, 1613 (2003);  arXiv : hep-th/0112121.
\bibitem{reyes} C. ~M. ~Reyes, Physical Review D \textbf{80}, 105008 (2009).


\end{thebibliography}
\end{document}